# Localization in an acoustic cavitation cloud

Miao Boya and An Yu
Department of Physics, Tsinghua University, Beijing 100084, People's Republic of China

Using a nonlinear sound wave equation for a bubbly liquid in conjunction with an equation for bubble pulsation, we predict and experimentally demonstrate the appearance of a gap in the frequency spectrum of a sound wave propagating in a cavitation cloud comprising bubbles. For bubbles with an ambient radius of 100 μm, the calculations revealed that this gap corresponds to the phenomenon of sound wave localization. For bubbles with an ambient radius of 120 μm, this spectral gap relates to a forbidden band of the sound wave. In the experiment, we observed the predicted gap in the frequency spectrum in soda water; however, in tap water, no spectral gap was present because the bubbles were much smaller than 100 μm.

When a high-amplitude ultrasonic wave passes through a liquid, many tiny bubbles are generated that violently and nonlinearly pulsate, undergoing periodic expansion and compression. These bubbles form a 'cavitation cloud', which can display various patterns and also emit faint glow (*i.e*., sonoluminescence) [1, 2]. Many cavitation phenomena arise through the nonlinear response of oscillating bubbles to the sound field. Here, we reveal a new phenomenon caused by the nonlinearity of the ultrasonic wave itself. Although this has not been predicted or observed previously in a cavitation cloud, this phenomenon may be related to a well-known quantum effect — Anderson localization [3]. We are able to connect a classical phenomenon to a quantum effect because the latter is intrinsically due to the wave properties [4]. Quantum effects are sometimes observed in classical ultrasonic or electromagnetic waves. The study of such phenomena contributes to our understanding of the associated quantum effects.

Since Anderson proposed the concept of electron localization in disordered solids, localization has been observed not only for electrons, but also for photons [5] and phonons [6]. Local resonance in certain media also can block the transmission of photons or phonons at particular frequencies [7], resulting in photon or phonon localization. Similarly, numerical simulation of sound wave propagation in a bubbly liquid predicted a forbidden band slightly above the Minnaert resonance frequency [8] because of multiple scattering by bubbles [9,10,11]. However, to date, there has been no report of the experimental observation of such a phenomenon.

In the present work, we theoretically predict frequency-dependent transmission of sound waves through a cavitation cloud, and then validate these predictions experimentally by observing cavitation clouds in tap and soda water. In a bubbly liquid such as soda water, the cavitation bubbles are usually much larger and the bubble number density is much higher than in regular tap water. Using mean field theory of an intense ultrasonic wave propagating in a bubbly liquid, under the assumption that all bubbles had an ambient radius of 100 μm, we numerically predicted a gap in the frequency spectrum. After trying many different liquids, we observed the predicted phenomenon in soda water.

The experimental set-up is simple, comprising a cylindrical ultrasonic horn tip with diameter of 2 cm (UH-800A, Autoscience Instrument Co. Ltd.) that was slightly immersed into liquid. A hydrophone (CS-5) was placed in the liquid at a fixed distance from the horn tip to measure the acoustic signals, which were displayed on a Tektronix DPO 2024 oscilloscope. Although the frequency of the ultrasonic wave was 20 kHz, nonlinear bubble oscillations in the cavitation cloud generated a broadband frequency signal [12], equivalent to the range of the frequency sweep used in the present experiment. Through simple analysis of the frequency spectrum, we gained an understanding of the frequency dependence of sound wave transmission in the cavitation cloud. The fast Fourier transforms (FFT) of the measured sound wave signals are shown in FIG. 1(a,b) for tap water and soda water, respectively. In soda water, although the ultrasound intensity was still high, it was reduced relative to that in tap water to ensure the cavitation cloud was confined to the region surrounding the horn tip and to avoid excessive bubble nucleation throughout the liquid. We observed a spectral gap in the low-frequency region for soda water between 40 and 200 kHz (inset in FIG. 1(b)), but no gap was observed for tap water. Thus, sound waves with frequencies in this range are not transmitted through the cavitation cloud.

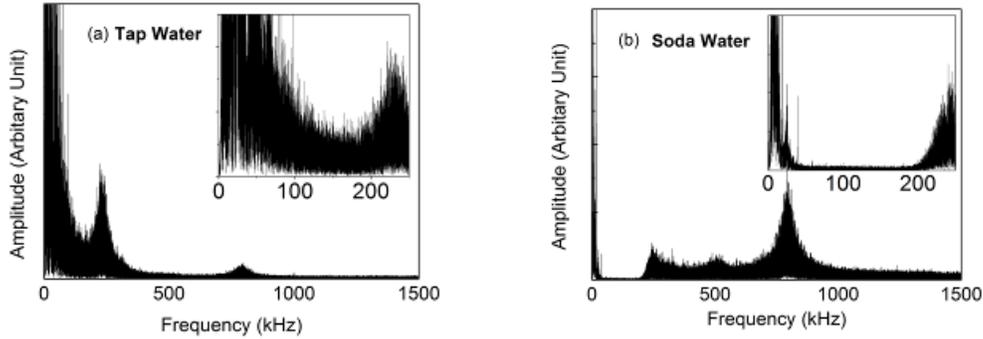

FIG. 1. Measured frequency spectrum of cavitation clouds in (a) tap water and (b) soda water. The spectrum was obtained by FFT of the time-domain signal of 200 acoustic cycles at the edge of the cavitation cloud. For soda water, a large spectral gap is present from 40–200 kHz. The peaks around 250 and 750 kHz are perturbation signals due to the hydrophone's resonant response. The sensitivity of the hydrophone decreases at high frequency and no correction was made for this in the present measurement.

To further investigate the differences in the frequency spectrum between tap and soda water, we imaged the cavitation cloud immediately after switching off the ultrasound source using a digital microscope (3R-MSTVUSB273, Anyty). During cavitation, bubbles pulsate violently and move chaotically. When the sound source is switched off, the bubbles exist in a transient steady state at their ambient size. From many images, we estimated the ambient bubble radius. As shown in FIG 2, the average ambient radii in soda water and tap water were 0.10 and 0.020 mm, respectively. The larger bubble size and bubble number density in soda water indicates that the spectral gap is unlikely to be predicted [9, 10, 11] using multiple scattering theory. In multiple scattering theory, neither the intensity of the ultrasonic field nor the bubble size are important. However, the conditions of the experiment are close to those set in our numerical calculation. Moreover, the experimentally observed results are similar to what we predicted by numerically solving the nonlinear sound wave equation in the cavitation cloud.

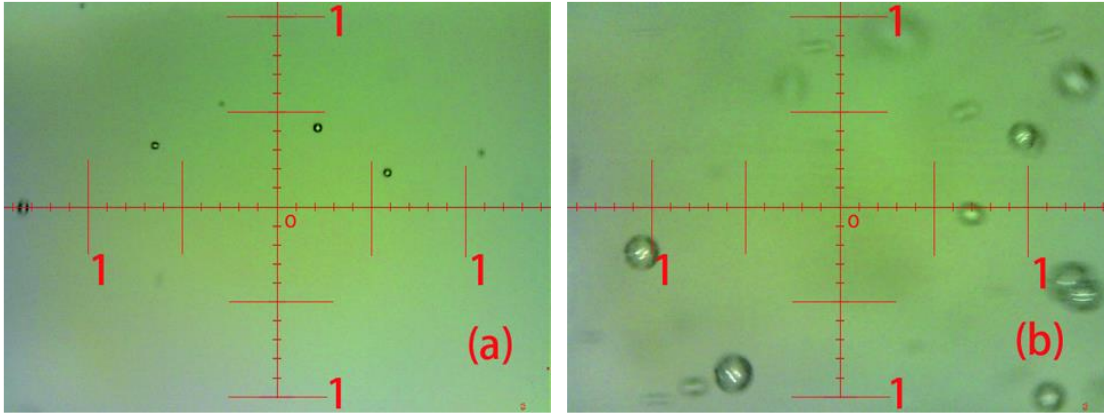

FIG. 2. Microscopic snapshots in (a) tap water and (b) soda water captured by a portable digital microscope. The scale units are millimeters. The ambient bubble radius in tap water and soda water were about 0.020 and 0.10 mm, respectively. In addition, the bubble number density in soda water was higher than in tap water.

The nonlinear sound wave equation in a bubbly liquid was proposed by Zabolotskaya and Soluyan [13]. Using this equation in conjunction with the equation for bubble pulsation (*i.e.*, the Rayleigh–Plesset equation), we can describe how a high-amplitude ultrasonic wave produces a cavitation cloud from a bubbly liquid, and how the sound wave propagates through the cloud [12]. For simplicity, we assumed that all bubbles are identical and homogenously distributed in the liquid. Because the horn tip (*i.e.*, the ultrasonic source) is cylindrical, we used an axisymmetric approximation to simplify the calculation. The driving frequency and geometric dimensions of the sound source were the same as those used in the experiment. According to the numerical simulation, under the condition that the number density of bubbles is about $1 \times 10^9$ m$^{-3}$, the gap in the frequency spectrum is only pronounced when the ambient bubble radius is about 80–120 µm. If the ambient bubble radius is smaller than 80 µm, a less pronounced frequency gap might be observed in experiments. Thus, the smaller the bubble size, the less clear the gap in the frequency spectrum. For ambient radii greater than 120 µm, as the bubble size increases, the gas fraction increases and the bubbly liquid becomes closer to a gas phase. As a consequence, all frequencies were blocked by the cavitation cloud with the blocking effect becoming more pronounced as the bubble size increased. A typical frequency spectrum for an ambient radius of 20 µm is shown in FIG. 3(a), which may represent the case of tap water. No gap was observed, similar to the experimental observation. If we assume the ambient bubble radius is 100 µm and the number density of bubbles ($N$) is $1 \times 10^9$ m$^{-3}$ (FIG. 3(b)), a gap appears in approximately the same frequency range (*i.e.*, 40–200 kHz) as the experimental observation shown in FIG. 1(b). The Minnaert resonance frequency, which can be calculated by the formula $f = \dfrac{1}{2\pi a}\left(\dfrac{3\gamma p_0}{\rho}\right)^{1/2}$, is about 30 kHz for an ambient radius of 100 µm. Thus, the gap we observed is different from that predicted by multiple scattering theory. By contrast, the nonlinear sound wave equation considers the secondary radiation of bubbles, which is different from multiple scattering.

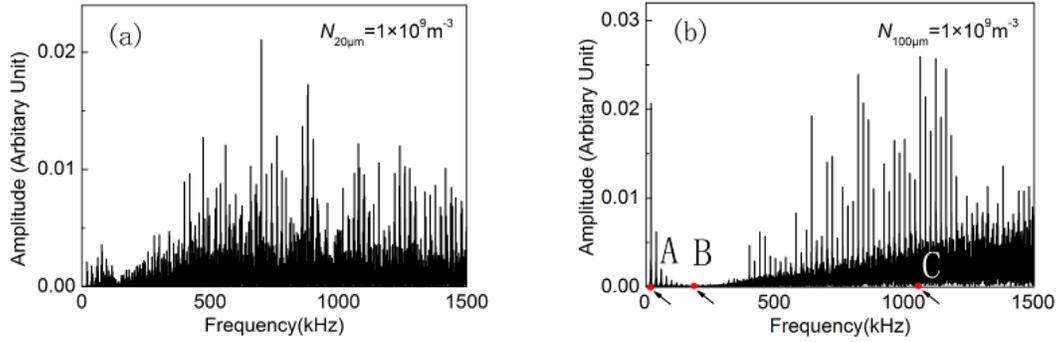

FIG. 3. Calculated frequency spectrum 31 mm from the surface of the horn tip at the edge of the cavitation cloud. The bubble number density ($N$) was $1 \times 10^9$ m$^{-3}$ and the ambient bubble radius was (a) 20 µm and (b) 100 µm. The data of 200 acoustic cycles were used.

In general, a spectral gap in the low-frequency region signifies the occurrence of wave localization [4]. To investigate the frequency dependence of ultrasonic wave transmission, we selected three different frequencies that are labeled A, B, and C in FIG. 3(b), for which we calculated the spatial distribution of the sound pressure along the symmetrical axis of the horn tip. In the calculation, the thickness of the cavitation cloud was assumed to be 30 mm. Point A (20 kHz) is the fundamental frequency of the ultrasonic source, and the sound pressure corresponding to this frequency as a function of the distance to the horn tip surface is shown in FIG. 4(a). Most of the acoustic energy at 20 kHz is confined within the cavitation cloud. At distances greater than 30 mm (*i.e.*, outside the cavitation cloud), the sound pressure decreases, meaning that some ultrasonic waves at 20 kHz are transmitted through the cavitation cloud. The sound pressure as a function of the distance from the surface of the horn tip corresponding to point B in FIG. 3(b) is shown in FIG. 4 (b). Point B corresponds to 180 kHz, which is within the spectral gap. At this frequency, the sound wave is localized within the cavitation cloud and is not transmitted. Point C is in the region of broadband noise. Sound waves at this frequency are transmitted through the cavitation cloud with little attenuation (FIG. 3(c)). Thus, the gap in the frequency spectrum corresponds to the localization of the sound wave. However, this argument is valid for 100-µm bubbles, but not for the larger (120 µm) bubbles investigated.

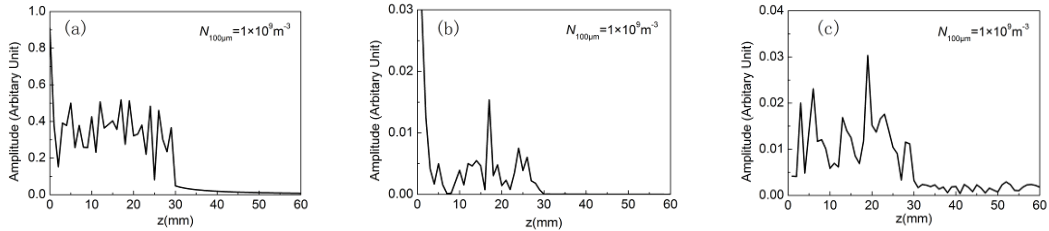

**FIG. 4.** Spatial distribution of the sound pressure along the symmetrical axis of the horn tip corresponding to the three points labeled in FIG. 3(b). (a) Point A, 20 kHz, (b) point B, 180 kHz, and (c) point C, 1050 kHz.

For an ambient radius of 120 µm, a spectral gap can still be observed, as shown in FIG. 5(a). For point A in FIG. 5(a), which is within the spectral gap, the sound pressure as a function of distance from the surface of the horn tip is shown in FIG. 5(b). Numerical fitting of the curve reveals that the sound pressure decays exponentially. Thus, sound waves at frequencies within the spectral gap cannot propagate in the cavitation cloud. This also means that the spectral gap observed for 120-μm bubbles with a number density of $1 \times 10^9$ $m^{-3}$ probably corresponds to a forbidden band of the sound wave. The question therefore arises whether the spectral gap observed in FIG. 1(b) is due to the localization described in FIG. 4 (b) or to the sound wave being blocked by the cavitation cloud, as was the case in FIG. 5(b). In reality, cavitation clouds comprise bubbles with different ambient radii. For simplicity, we performed the calculations for a mixture of 95% 100-μm bubbles and 5% 120-μm bubbles, the results of which are shown in FIG. 6. The frequency spectrum is very similar to that of 120-μm bubbles (FIG. 5(a)). However, for the point in the spectral gap, the spatial distribution of the sound pressure along the symmetrical axis of the horn tip no longer decays exponentially; rather, it fluctuates, which reflects wave localization. This mixed bubble case may more closely represent the experimental conditions than the case with bubbles of a single size. If the calculations were performed for bubbles with a distribution of sizes, the calculated result should match the experimental observation even more closely.

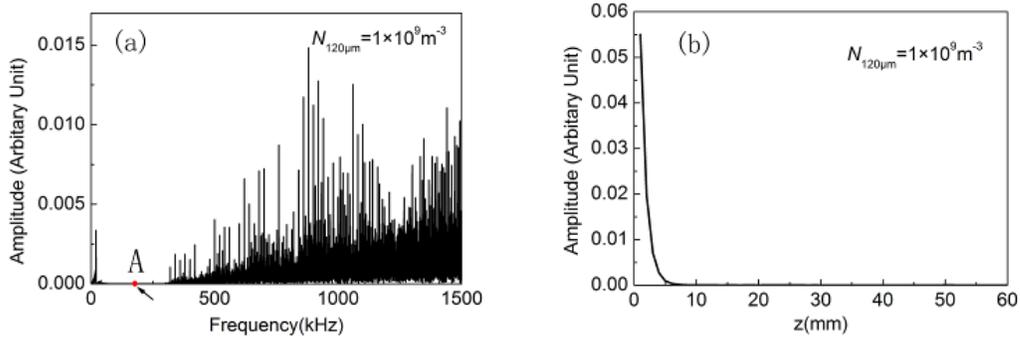

**FIG. 5.** (a) Calculated frequency spectrum 31 mm from the surface of the horn tip at the edge of the cavitation cloud. In the calculation, the bubble number density ($N$) was $1 \times 10^9$ $m^{-3}$ and the ambient bubble radius was 120 μm. The data for 200 ultrasonic cycles were used. (b) Spatial distribution of the sound pressure along the symmetrical axis of the horn tip corresponding to point A in panel (a).

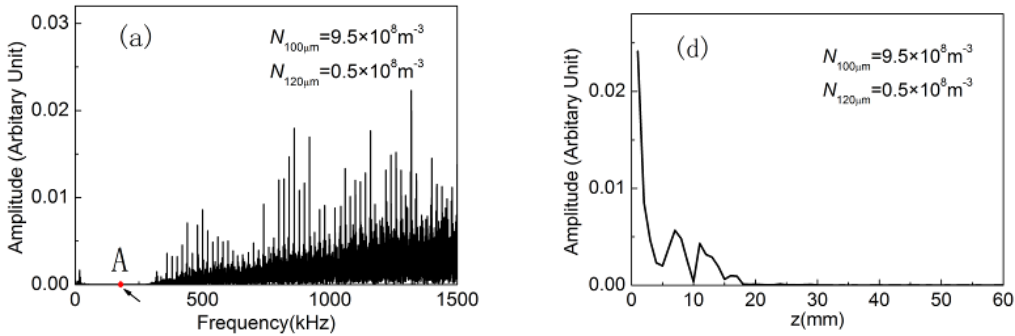

FIG. 6. (a) Calculated frequency spectrum 31 mm from the surface of the horn tip surface at the edge of the cavitation cloud. In the calculation, we considered bubbles with different ambient radii: $R_0 = 100$ μm with number density ($N_{100\ \mu m}$) of $9.5 \times 10^8$ $m^{-3}$ and $R_0 = 120$ μm with $N_{120\ \mu m} = 0.5 \times 10^8$ $m^{-3}$. The data of 200 ultrasonic cycles were used. (b) Spatial

distribution of the sound pressure along the symmetrical axis of the horn tip corresponding to point A in panel (a).

In summary, we employed a theoretical method based on a nonlinear ultrasonic wave equation and a bubble dynamics equation to simulate sound wave propagation in a cavitation cloud. The calculations predicted a gap in the frequency spectrum for bubbles with an ambient radius of about 100 μm. Experimentally, in soda water, we observed the predicted phenomenon. Further calculations revealed that for a cavitation cloud comprising 100-μm bubbles, the spectral gap is due to sound wave localization. By contrast, for 120-μm bubbles, the spectral gap is attributed to forbidden bands that prevent sound wave propagation. In the experiment, it is likely that the bubbles had a wide distribution of ambient radii and that the spectral gap resulted from sound wave localization. In contrast to that in soda water, the cavitation cloud in tap water consists of bubbles much smaller than 100 μm; therefore, in tap water, no spectral gap was observed.

The authors are grateful to Prof. L. M. Li for helpful discussions on wave localization.

This work is supported by the NSFC under Grant No. 11334005 and the Specialized Research Fund for the Doctoral Program of Higher Education of China under Grant No. 20120002110031.

[1] H. Frenzel, and H. Schultes, Z. Phys. Chem. Abt. B **27B**, 421(1934).

[2] C. D. Ohl, T. Kurz, R. Geisler, O. Lindau and W. Lauterborn, Phil. Trans. R. Soc. Lond. A**357**, 269(1999).

[3] P. W. Anderson, Phys. Rev. **109**, 1492 (1958).

[4] T Brandes and S. Kettemann, *Anderson Localization and Its Ramifications: Disorder, Phase Coherence, and Electron Correlations* (Springer, 2003), p.5.

[5] T. J. Shepherd, J. Mod. Optic. **3**, 657 (1994).

[6] E. Yablonovitch, Phys. Rev. Lett. **58**, 2059 (1987).

[7] Z. Y. Liu, X. X. Zhang, Y. W. Mao, et al., Science **289**, 1734 (2000).

[8] M. Minnaert, Philos. Mag. **16**, 235 (1933).

[9] Z. Ye and A. Alvarez, Phys. Rev. Lett. **80**, 3503 (1998).

[10] M. Kafesaki, R. S. Penciu, and E. N. Economou, Phys. Rev. Lett. **84**, 6050 (2000).

[11] B. Liang, and J. C. Cheng, Phys. Rev. E **75**, 016605 (2007).

[12] Y. An, Phys. Rev. E **85**，016305 (2012).

[13] E.A. Zabolotskaya, S.I. Soluyan, Sov. Phys. Acoust. **18**, 396 (1973).